\documentclass{elsart}

\usepackage{epsfig}

\begin{document}
\begin{frontmatter}
\title{A random walker on a ratchet}

\author{Jos\'e L. Mateos\thanksref{EMAIL}}
\address{Instituto de F\'{\i}sica, 
Universidad Nacional Aut\'onoma de M\'exico, \\
Apartado Postal 20-364, 01000 M\'exico, D.F., M\'exico}
\thanks[EMAIL]{E-mail: mateos@fisica.unam.mx; \\
Fax: (52) (55) 5622 5015; Phone: (52) (55) 5622 5130}

\begin{abstract}

We analyze a model for a walker moving on a ratchet potential. 
This model is motivated by the properties of transport of 
motor proteins, like kinesin and myosin. The walker consists of two 
feet represented as two particles coupled nonlinearly through a bistable potential. 
In contrast to linear coupling, the bistable potential admits a richer dynamics where 
the ordering of the particles can alternate during the walking. The transitions between the two 
stable states on the bistable potential correspond to a walking with alternating 
particles. We distinguish between two main walking styles: alternating and no alternating, 
resembling the hand-over-hand and the inchworm walking in motor proteins, respectively.   
When the equilibrium distance between the two particles divided by the periodicity
of the ratchet is an integer, we obtain a maximum for the current, indicating optimal transport.                           
\end{abstract}

\end{frontmatter}

\section{Introduction}

In recent years, advances in non-equilibrium statistical physics have revealed various instances 
of the surprising phenomenon of noise enhanced order, such as stochastic resonance 
\cite{sr1,sr2,sr3}, Brownian motors, ratchets or noise-induced transport 
\cite{rev1,rev2,rev3}. These remarkable phenomena occur due to the 
constructive role of noise in nonlinear dynamical systems \cite{shura}.  
Noise-induced, directed transport in a spatially periodic system in thermal 
equilibrium is ruled out by the second law of thermodynamics. Therefore, 
in order to generate transport, the system has to be driven away from 
thermal equilibrium by an additional deterministic or stochastic
force. In the most interesting situation, these forces are unbiased, 
that is, their temporal, spatial or ensemble averages  
vanish. Besides the breaking of thermal equilibrium, another 
important requirement to get directed transport in a spatially periodic system 
is the breaking of the spatial inversion symmetry. We speak then of Brownian 
motors, ratchet potentials, or, in the biological realm, of molecular motors. 
This recent burst of work is motivated in part by the challenge to explain the unidirectional 
transport of molecular motors in the biological realm \cite{vale,kel,how}. 
    
One particular motor protein, kinesin, has attracted considerable attention, 
motivated by experimental results in which the dynamical details of its motion 
can be measured \cite{vale,how}. Kinesin is a protein with two heads that perform 
a walking on microtubules inside cells. Motivated by these experimental results, 
several authors \cite{der,stra,klum,els,bier,kan,dan,wang} have introduced diverse models in 
order to understand the particular walking of kinesin. Usually, these models consider 
two coupled particles on a ratchet potential that represents the periodic asymmetric 
structure of microtubules. In these papers the authors consider  
a linear elastic coupling between the particles. This coupling
implies that the order of the particles cannot change. However, in recent experiments,
it was found that kinesin moves processively alternating its two heads in a way called
hand over hand \cite{blo3,alvaro,selvin2}. Other processive motor proteins 
that move on actin filaments in a hand-over-hand way are myosin V \cite{selvin1,selvin3} 
and myosin VI \cite{selvin4,spudich}. 

We introduced in this paper a model inspired by the walking of motor proteins, like
kinesin on microtubules, but that is not restricted to walking of motor proteins. 
The model can be useful to describe as well the walking of macroscopic objects in 
the presence of fluctuations. It consists of two particles coupled through a 
nonlinear bistable potential and subjected to independent white noises. This system 
of two coupled particles is acted upon by a spatially-periodic force, due to the 
presence of a ratchet potential and, additionally, we have a common time-dependent periodic 
force. We are interested in analyzing the trayectories of the walker
following in detail the motion of each of the particles, and we are also interested in
the current or noise-induced transport for this system. 
 
\section{The model of a walker with two Brownian motors}

The model considers a walker moving on an asymmetric ratchet potential 
mediated by noise. This walker has two feet that are represented as two 
particles coupled nonlinearly through a bistable potential \cite{upon,spie,matfnl}. 
The walker moves along a track formed by an asymmetric ratchet potential, 
and is subjected to the influence of two independent white noises acting on 
the two particles and a common external harmonic force. The stochastic 
differential equations for the two particles, represented by $x$ and $y$, 
in the overdamped regime, are:

\begin{equation}
m\gamma \dot{x} = -\partial_{x}V(x) - \partial_{x}V_{b}(x-y) + m\gamma\sqrt{2D}\xi_{1}(t) + F_{D}\sin(\Omega t+\varphi)   
\end{equation}

\begin{equation}
m\gamma \dot{y} = -\partial_{y}V(y) - \partial_{y}V_{b}(x-y) + m\gamma\sqrt{2D}\xi_{2}(t) + F_{D}\sin(\Omega t+\varphi).
\end{equation}

\noindent where $m$ is the mass of each particle, $\gamma$ is the friction coefficient, 
$-\partial_{x}V(x)$ is the force due to the ratchet potential, $-\partial_{x}V_{b}(x-y)$ is the coupling force due to 
the bistable potential. The common external harmonic force has three parameters: the amplitude force $F_{D}$, 
the frequency $\Omega$ and the initial phase $\varphi$.

\noindent These equations represent two coupled particles on a periodic asymmetric
ratchet potential given by \cite{matprl}

\begin{equation}
V(x) = V_1 - V_{R} \left [\sin {\frac{2\pi (x-x_0)}{L}} - {\frac{1}{4}} \sin {\frac{4\pi (x-x_0)}{L}} \right ].
\end{equation}

\noindent where $L$ is the period of the potential $V(x + L) = V(x)$; the other constants
will be discussed later. Additionally, these particles are coupled by the {\em nonlinear} cubic force 
coming from a bistable potential $V_{b}(x - y)$ given by

 \begin{equation}
V_{b}(x - y) = V_{b} + V_{b} \left[ {\frac{(x - y)^4}{l^4}} - 2{\frac{(x - y)^2}{l^2}} \right].
\end{equation}

\noindent Here, $V_{b}$ is the amplitude of the bistable potential and represents
the coupling strenght between the particles, and $2l$ is the distance between the two 
minima. 

Finally, the parameter $D$ is the intensity of the zero-mean statistically independent Gaussian 
white noises $\xi_{1}(t)$ and $\xi_{2}(t)$  acting on particles $x$ and $y$, 
respectively. Being statistically independent, the following equation is satisfied:

\begin{equation}
\langle \xi_{i}(t) \xi_{j}(s) \rangle = \delta_{ij} \delta (t - s).
\end{equation}

Let us derive now dimensionless equations of motion for the model. We use as the 
characteristic lenght scale the period of the ratchet potential $L$, the characteristic time
scale will be given by the inverse of the friction coeficient $\tau = 1/ \gamma$, and the
charactersitic force is $mL\gamma^{2 }$. 
Let us define the following dimensionless units: $x^{\prime } = x/L$,
$x_{0}^{\prime } = x_{0}/L$, $y^{\prime } = y/L$,
$y_{0}^{\prime } = y_{0}/L$, $t^{\prime } = \gamma t$, ${l^{\prime } = l/L}$,
$\Omega^{\prime} = \Omega/\gamma$ and $D^{\prime} = D/ \gamma L^{2}$. 
The dimensionless equations of motion, after renaming the variables again without 
the primes, become 

\vfill\eject

\begin{equation}
\dot{x} = -\partial_{x}V(x) -\partial_{x} V_{b}(x - y) + \sqrt{2D}\xi_{1}(t) + A \sin(\Omega t + \varphi),
\end{equation}

\begin{equation}
\dot{y} = -\partial_{y}V(y) - \partial_{y}V_{b}(x - y) + \sqrt{2D}\xi_{2}(t) + A \sin(\Omega t + \varphi),
\end{equation}

The dimensionless ratchet potential is

\begin{equation}
V(x) = C -  U_{R} \left [\sin 2\pi (x-x_{0}) + {\frac{1}{4}} \sin 4\pi (x-x_{0}) \right ].
\end{equation}

\noindent The constant $C = -U_{R} (\sin 2\pi x_{0} + 0.25 \sin 4\pi x_{0})$ is such 
that $V(0)=0$. The constant $x_{0}$ is introduced in order to center the minima of the periodic 
potential on the integers \cite{matprl}. 

The dimensionless bistable potential is given by

 \begin{equation}
V_{b}(x - y) = U_{b} \left[1 + {\frac{(x - y)^4}{l^4}} - 2{\frac{(x - y)^2}{l^2}} \right].
\end{equation}

Here the dimensionless amplitude of the ratchet potential is given by  
$U_{R} = V_{R}/(mL^{2}\gamma^{2})$, and the dimensionless amplitude of the bistable potential 
$U_{b} = V_{b}/(mL^{2}\gamma^{2})$. The amplitude of the external force is $A = F_{D}/(mL\gamma^{2})$.

In Fig. 1 we depict the ratchet (solid line) and the bistable (dashed line) potential for this model.
The dotted line is the sum of both potentials. As is clear in the figure, the total potential has three
minima instead of two, due to the interplay between the stable points in both potentials. These means
that we have three stable equilibria configurations for the walker: $-l$, $0$ and $l$. In our model the
situation is even more complicated since we have two particles in a potential like in Fig. 1. We can
think of this problem as a single particle in a two-dimensional potential given by
$\Psi(x,y) = V(x) + V(y) + V_{b}(x - y)$. A discussion of this 2D problem and the analysis of its
bifurcations can be seen in \cite{upon,spie,matfnl}.   

\begin{figure} 
\begin{center} 
\epsfig{figure=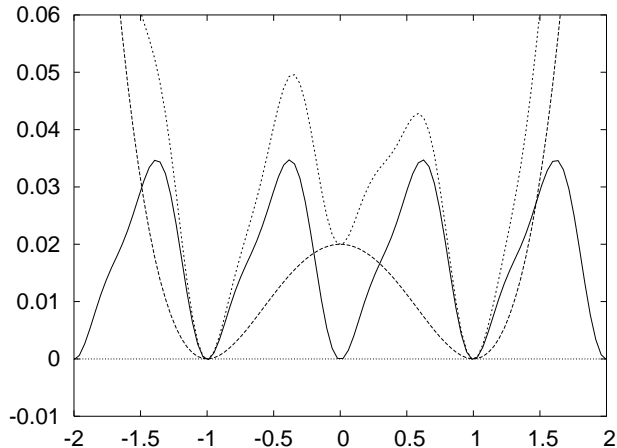,width=9.0cm} 
\end{center} 
\caption{The ratchet (solid line) and bistable potential (dashed line) 
for the random walker model in the case $l = 1$. The dotted line is the sum of both potentials.}
\label{fig1} 
\end{figure}

This model is different from previous ones described in the literature, since it 
incorporates a {\em nonlinear} coupling between the two particles through the bistable 
potential, as has been discussed before \cite{upon,spie,matfnl}. It is important to 
stress the following: the coupling through the bistable potential involves the variable
$x - y$. This variable can be positive, negative or zero. When $x - y > 0$ 
the $x$ particle is ahead of the $y$ particle. On the other hand, when
$x - y < 0$ the $y$ particle is ahead of the $x$ particle.
Therefore, the transitions between the two stable states in the bistable potential
correspond to an exchange of the order between the particles. The minima, 
corresponding to the two stable points in the bistable potential, are located at 
$x - y = l$ and $x - y = -l$, that is, when the distance between the two particles is $l$.
Thus, we have two equilibrium configurations for the walker: $x - y = l$ and $x > y$, or
$x - y = -l$ and $x < y$. The local maximum at the origin in the bistable potential is unstable 
without the ratchet, but it can become a third stable configuration in the presence of the
ratchet potential, as can be seen in Fig. 1. This stable state 
corresponds to the case when $x - y = 0$, that is, when the two particles coincide in space.     
So, we can think of a state oscillating in the bistable potential back and forth between 
the two minima, as the walking of a motor protein alternating its two heads, or a walker 
alternating its two feet. This nonlinear coupling allow us then to consider a 
very important aspect of a real walking that was lacking in previous models: the possibility 
of alternating the two feet. In the models given in \cite{der,stra,klum,els,dan,wang}, the coupling between 
the particles is linear (a harmonic spring) and thus the particles cannot alternate positions. 
They simply can approach to each other, but once you have an ordering, say $x>y$, 
the ordering remains for the rest of the dynamics. In our case, on the other hand, we can 
have several types of walking: alternating random walking, where the two particles 
alternate their order randomly when the walker moves through the ratchet 
(hand over hand) or; rigid random walking, where the two particles move on the ratchet 
without exchanging their order (inchworm). 

\begin{figure} 
\begin{center} 
\epsfig{figure=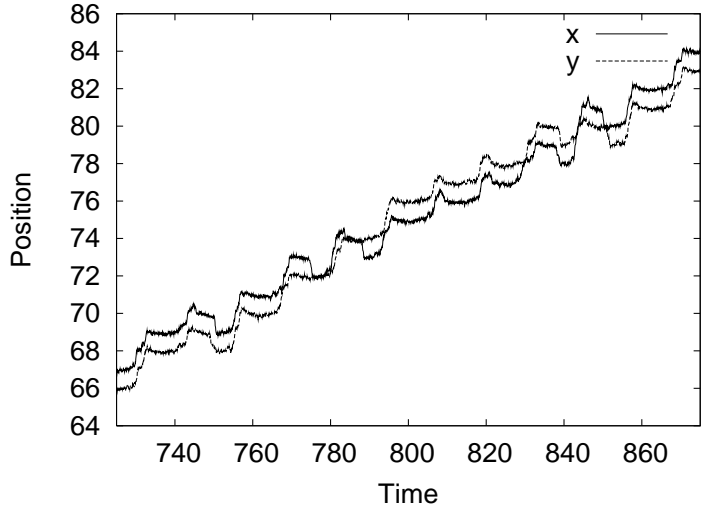,width=10.0cm} 
\end{center} 
\caption{A typical trajectory of a walker moving along the ratchet.
The solid line represents the $x$ particle and the dashed line 
the $y$ particle. The parameters are: $A = 1.0$, $D = 0.1$,
$\Omega = 0.5$, $l = 1.0$ and $r = 0.78$.}
\label{fig2} 
\end{figure}

\begin{figure} 
\begin{center} 
\epsfig{figure=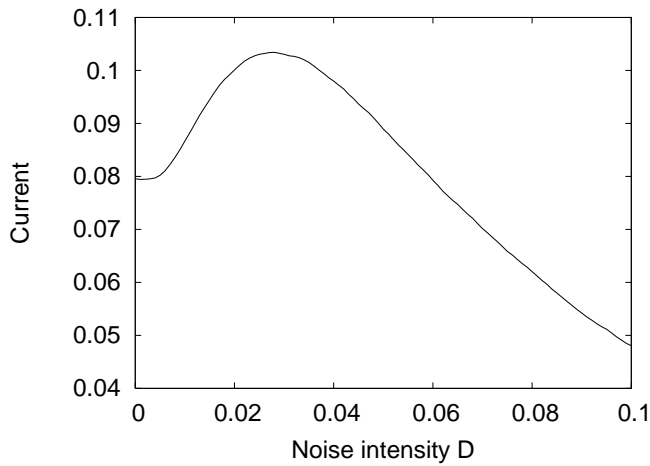,width=9.0cm} 
\end{center} \caption{The current as a function of noise intensity $D$. The parameters
are: $A = 1.0$, $\Omega = 0.5$, $l = 1.0$ and $r = 1.57$.}
\label{fig3} 
\end{figure}

Coming back to the stochastic equations that define our model, we notice that an
important parameter is $r = U_{b}/U_{R}$, the ratio of two barrier heights: 
the bistable and the ratchet. Both are energetic barriers that have to be overcome in 
order to perform a particular walking: alternating or rigid. 
If we overcome only the ratchet barrier, then we have a rigid walking, whereas if the walker
can overcome both barriers, then the walking can alternate the feet. 
From this particular coupling we see a clear connection between the phenomena of
stochastic resonance and Brownian motors. One expects that in an optimal situation,
aided by the stochastic resonance mechanism, one can transit very efficiently between
the two states in the bistable potential and at the same time walk optimally on the ratchet
potential using an alteration of the two particles. This might happen when the equilibrium
distance $l$ between particles coincide with the periodicity of the ratchet potential. In 
our units this case corresponds to $l = 1$.

\section{Numerical results}

In this section we will solve numerically the Langevin equations of motion to obtain the
trajectories of the walker. We use a stochastic fourth-order Runge-Kutta algorithm to 
solve the system of stochastic differential equations with additive Gaussian white noise. 
This means that we use a fourth-order Runge-Kutta algorithm for the deterministic part and a 
random number generator to obtain Gaussian-distributed random numbers from
uniformly distributed ones in the unit interval, using the Box-Mueller algorithm, as described 
in \cite{press}. We calculate the current, which is simply the ensemble average velocity 
of the center of mass of the walker, where the center of mass is given by $z(t) = (x(t) + y(t))/2$. 
In the numerical results that we present now, we fix the value of $\Omega = 0.5$. 
The other parameters are indicated in the text or in the figure captions. 
The quantities that we compute involve an averaging of an ensemble with different 
initial random phases $\varphi$ uniformly distributed on a circle between $0$ and $2\pi$. 

\begin{figure} 
\begin{center} 
\epsfig{figure=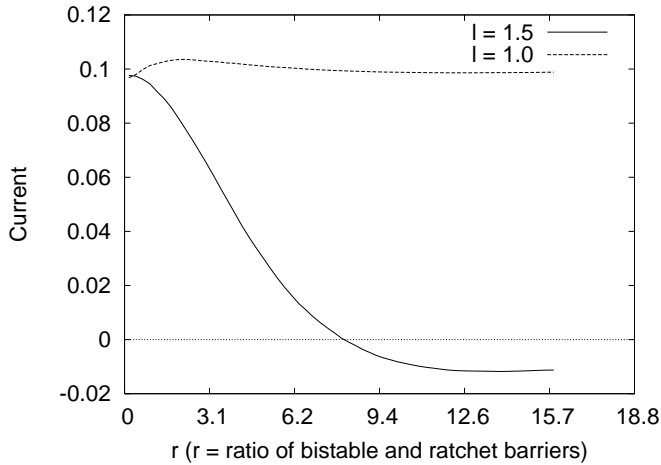,width=9.0cm} 
\end{center} 
\caption{The current as a function of the ratio of the amplitudes of the 
ratchet and bistable potential barriers $r$. The dashed line corresponds to $l = 1$ and
the solid line to $l = 1.5$. Notice the current reversal for the latter case.
The parameters are: $A = 1.0$, $D = 0.03$ and $\Omega = 0.5$.}
\label{fig4} 
\end{figure}

\begin{figure} 
\begin{center} 
\epsfig{figure=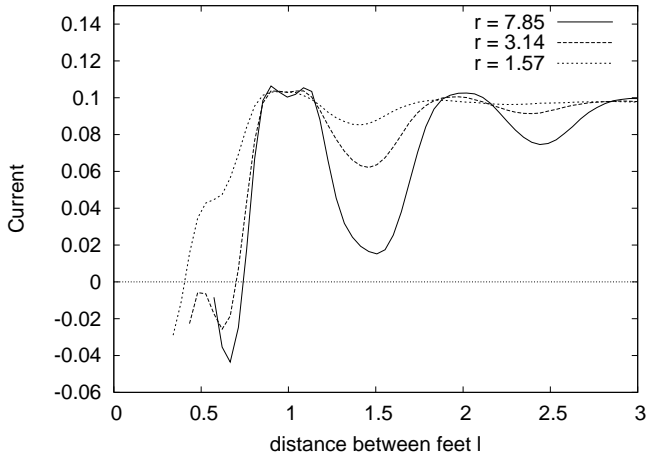,width=9.0cm} 
\end{center} 
\caption{The current as a function of the equilibrium distance between particles $l$.
The different curves correspond to different values of the ratio $r$. 
The parameters are: $A = 1.0$, $D = 0.03$ and $\Omega = 0.5$.}
\label{fig5} 
\end{figure}

In Fig. 2 we show a typical trajectory for the walker. The solid line corresponds to the
$x$ particle and the dashed line to the $y$ particle. Notice that due to the coupling the
two particles tend to move together and the center of mass advance with a positive current.
As can be seen in the figure, the two feet can exchange positions and different types of
walking patterns can be observed. For instance, the walker can overcome the ratchet
barriers without exchanging the order of the particles for a while, but later on the feet
tend to be together and the walker jumps in this way to the next minimum 
in the ratchet and finally the order of the feet can change.
In Fig 3 we show the current as a function of noise intensity. We notice that the current
increases until it reaches a maximum and then tend to decrease. This remind us the phenomenon
of stochastic resonance. For zero noise the current is finite, since we have a large amplitude $A = 1$
for the external forcing. So we obtain a nonzero current even in the deterministic limit.
In Fig. 4 we depict the current as a function of $r$, the ratio between the bistable and ratchet 
potentials, for two values of the equilibrium distance between feet $l$. For $l = 1$ we see that
the current is almost constant, but for $l = 1.5$ the current depends strongly on $r$ and can even
reverse sign. So we obtain a current reversal as a function of the ratio $r$.
Finally, in Fig. 5 we show the current as a function of the equilibrium distance between feet $l$, for
different values of the ratio $r$. We did not show the current for small values of $l$, since in this case
the bistable potential can be very large in comparison with the other terms in the Langevin equations.   
We notice that the current attains a maximum when $l$ is near an integer. In some case the 
maximum is not exactly at the integers, but close to it. In fact, for large values of
the ratio $r$ (for instance $r = 7.85$ in Fig. 5), the current develops a local minimum at $l = 1$. 
We are planning to analyze further this effect for other parameters in order to determine 
its possible generality. This means that the walker can move through the ratchet in a very 
efficient way when each particle is located close to a minimum of the ratchet. On the other hand, 
if the distance $l$ is in between integers, the current attains a minimum, indicating that 
the walker is unable to move efficiently.   

\section{Concluding remarks}

In summary, we have introduced a model for a random walker that consists of two particles 
coupled nonlinearly through a bistable potential, steeping on an asymmetric periodic 
ratchet potential. In contrast to linear coupling, the bistable potential admits a richer 
dynamics where the ordering of the particles can alternate. The dynamics then include 
two typical stepping patterns: alternating (hand over hand) and non alternating walking 
(inchworm). In our model we can obtain both types of walking, depending on the ratio
between the ratchet and the bistable barriers, and on the ratio between
the equilibrium distance of the particles and the periodicity of the ratchet. 
It is worth mentioning that according to recent experiments in motor proteins 
\cite{blo3,alvaro,selvin2,selvin1,selvin3,selvin4,spudich} the hand over hand type of walking
is more likely. We have calculated the current, defined as the average velocity of the 
center of mass, as a function of parameters related with the coupling and the distance 
between particles. In the case where the equilibrium distance between particles is a multiple 
of the periodicity of the ratchet potential, we obtained a maximum value for the current. 
Thus, we have a new model that may shed light on a number of currently interesting problems, 
ranging from noisy locomotion to transport of motor proteins, that establish a connection 
between Brownian motors and stochastic resonance.

\bigskip
{\bf Acknowledgements}
\medskip

The author gratefully acknowledges helpful discussions with Alexander Neiman, 
Frank Moss, Lutz Schimansky-Geier, Jan Freund, Igor Sokolov and Peter H\"anggi. 
Financial support from UNAM through project DGAPA-IN-111000, is acknowledged.
The author also wants to thank the Alexander von Humboldt Foundation for support.

\end{document}